\documentstyle[12pt,twoside,fleqn,espcrc1,epsfig]{article}

\newcommand{\beq}{\begin{equation}}
\newcommand{\eeq}{\end{equation}}
\newcommand{\bea}{\begin{eqnarray}}
\newcommand{\eea}{\end{eqnarray}}
\newcommand{\bce}{\begin{center}}
\newcommand{\ece}{\end{center}}
\newcommand{\etal}{{\it et al.}}

\newcommand{\eg}{{\it e.g.}}
\newcommand{\ie}{{\it i.e.}}
\newcommand{\tave}[1]{\langle\!\langle{#1}\rangle\!\rangle}

\newcommand{\AmS}{{\protect\the\textfont2
  A\kern-.1667em\lower.5ex\hbox{M}\kern-.125emS}}

\title{Duality and Chiral Restoration from Low-Mass Dileptons
 at the CERN-SpS} 

\author{Ralf Rapp\thanks{supported by the A.-v-.Humboldt Foundation as a
Feodor-Lynen fellow, and by US-DOE grant DE-FG02-88ER40388.} \\ 
\vspace{0.3cm}
Department of Physics and Astronomy, SUNY at Stony Brook, NY 11794-3800, USA}

\begin{document}

\maketitle

\begin{abstract}
We review recent theoretical progress in low-mass dilepton
production at CERN-SpS energies. Various hadronic approaches 
to calculate the vector correlator in hot/dense hadronic matter 
are discussed and confronted with each other. Possible 
consequences for the nature of chiral restoration are indicated.

\end{abstract}

\section{Introduction}

The spectra of dileptons as penetrating probes in ultra-relativistic 
heavy-ion collisions (URHIC's) are expected to provide important information 
on the properties of QCD under conditions of extreme temperature
and density, associated with the possible occurrence of the 
chiral/deconfinement phase transition(s).
Depending on the dilepton invariant mass region under consideration, various
signatures related to different properties of the strong interactions
may be studied. 

At high invariant masses, $M_{ll}\ge 3$~GeV, the interest is attached  
to the  heavy quark flavors charm and bottom and the experimental focus
is on the behavior of the heavy quarkonium bound states such as $J/\Psi$  
and $\Upsilon$. In a possible formed quark-gluon plasma the color
interaction of their constituents ($c\bar c$ or $b\bar b$)
is believed to be Debye-screened, eventually causing 
a dissolution of the bound state. Thus a depletion
of the $J/\Psi$ (or $\Upsilon$) resonance structures in the 
dilepton spectrum might signal the liberation of color charges, 
\ie, deconfinement. 
This important topic has been covered in several talks in the 
plenary~\cite{Satz,Cicalo} and parallel sessions~\cite{jpsi}. 

The intermediate mass region (IMR), which extends
from about 1~GeV up to the $c\bar c$ threshold at 3~GeV, has long
been proposed~\cite{Shu80} as the suitable window  to observe 
an increased yield of thermal radiation from an equilibrated
quark-gluon plasma through $q\bar q$ annihilation of the light  
 flavors $q=u,d,s$. On the one hand, hard processes such as 
Drell-Yan (DY) annihilation are already sufficiently suppressed, and, 
on the other hand, hadronic decay contributions are concentrated
at smaller masses; moreover, the sensitivity of the IMR
on temperature through thermal factors $\propto \exp(-M/T)$ 
strongly favors the contributions from early stages at high 
temperatures. The most important background besides DY pairs is due  
to associatedly produced $D$ and $\bar D$ mesons ('open charm'). 
An anomalously increased open charm contribution has 
been suggested to be the origin of the observed IMR enhancement in 
Pb+Pb collisions~\cite{bordalo}, but thermal radiation seems to be the
more natural explanation (we will briefly return to this issue at the 
end). 

Finally, the low-mass region (which is the main subject of this talk, 
see also Ref.~\cite{RW00} for a recent review)
is characterized by the non-perturbative physics of 
the light (constituent) quarks and their bound states building up  
the low-lying hadronic spectrum. The crucial feature that governs the 
strong interactions in  this energy regime is the (approximate) chiral 
symmetry of the QCD Lagrangian being spontaneously  broken in the ground
state of the theory as revealed by the formation of the chiral quark
condensate and the absence of equal-mass chiral partners among  
hadrons.  The occurence of a chiral phase transition restoring
the symmetry, as clearly evident from state-of-the-art lattice
calculations, thus necessarily implies a substantial reshaping 
of the light hadron spectrum. From dilepton observables in 
URHIC's one hopes to witness this through the direct decays of the
light vector mesons,  $\rho, \omega, \phi\to l^+l^-$. Here, owing to the
inherent time scales of 10-20 fm/c in heavy-ion reactions, 
the $\rho$ meson plays the by far dominant role as it has the 
shortest lifetime and the largest electromagnetic decay width.  
A significant emphasis in this talk will therefore be on the study
of $\rho$ mesons in hot and dense matter. 
One key question then is to what extent its medium modifications can be
related to chiral restoration, and, in particular, how the latter is realized 
in more general terms (\eg, do all masses $\to 0$, or do all widths 
$\to \infty$?). This inevitably necessitates a simultaneous 
treatment of the chiral partner of the $\rho$, the $a_1(1260)$, which,
unfortunately, is limited to the theoretical level as  
medium modifications in the axialvector channel are difficult to extract 
from experiment. To eventually arrive at reliable answers, the following 
two strategies are essential: 
(a) to impose model constraints within one's favorite approach, 
both theoretical (symmetries and related low-$T$/-$\mu_B$ theorems, 
QCD sum rules) 
and phenomenological (through independent experimental information);  
(b) to compare various approaches and their distinct (and common) 
features with each other. 
Both are included in the discussion of the microscopic hadronic
models for vector mesons/dilepton rates presented in Sect.~\ref{sec_models}, 
which is divided into a finite temperature and a finite density part. 
   
In Sect.~\ref{sec_dlspec} we will then proceed to the application
to dilepton spectra in URHIC's. This involves a further complication, 
namely the space-time description of the global reaction dynamics, 
which will be briefly addressed before comparing results to 
dilepton data from the SpS.

We end with some concluding remarks in Sect.~\ref{sec_concl}.  

\section{Electromagnetic Current Correlator and Thermal Dilepton Rates}
\label{sec_models}
The general form of thermal dilepton production rates from a  hot and 
dense medium can be decomposed as 
\beq
\frac{dN_{ll}}{d^4xd^4q}=L_{\mu\nu}(q) \ W^{\mu\nu}(M,\vec q ;\mu_B, T) 
\eeq
with the lepton tensor (for $m_l^2 \ll M^2 = q_0^2-\vec q^2$) 
\beq
L_{\mu\nu}(q)=-\frac{\alpha^2}{6\pi^3 M^2} \left( g_{\mu\nu} -
\frac{q_\mu q_\nu}{M^2} \right) \ . 
\eeq
The hadron tensor  $W^{\mu\nu}$ contains all the non-trivial information
on the hadronic medium of temperature $T$ and baryon chemical potential 
$\mu_B$. It is defined via
the thermal expectation value of the electromagnetic (e.m.) current-current 
correlator~\cite{MT85} 
\bea
W^{\mu\nu}(q) &=& -i \int d^4x \ e^{-iq\cdot x} \  
\tave{j^\mu_{\rm em}(x) j^\nu_{\rm em}(0)}_T  
\nonumber\\
 &=& \frac{-2}{\exp(q_0/T)-1} \ {\rm Im} \Pi^{\mu\nu}_{\rm em}(q) \ . 
\eea
Depending on the invariant masses probed, the e.m.~correlator can be 
described by either using hadronic degrees of freedom (saturated
by vector mesons within the well-established vector dominance model (VDM))
or the (perturbative) quark-antiquark vector correlator, \ie, 
\beq
{\rm Im} \Pi_{\rm em}^{\mu\nu}  = \left\{
\begin{array}{ll}
 \sum\limits_{V=\rho,\omega,\phi} ((m_V^{(0)})^2/g_V)^2 \ 
{\rm Im} D_V^{\mu\nu}(M,\vec q;\mu_B,T) & , \ M \le M_{dual}   
\vspace{0.3cm}
\\
(-g^{\mu\nu}+q^\mu q^\nu/M^2) \ (M^2/12\pi) \ N_c 
\sum\limits_{q=u,d,s} (e_q)^2  & , \ M \ge M_{dual} \ . 
\end{array}  \right. 
\label{ImPiem}
\eeq
In vacuum the transition region is located at a 'duality threshold'
of $M_{dual}\simeq 1.5$~GeV, as marked by the inverse process 
of $e^+e^-$ annihilation into hadrons. We will argue below
that medium effects in the correlator might be summarized as a lowering
of the duality threshold in hot and dense matter.  

In the remainder of this section we elucidate on various investigations  
that have been performed to study medium modifications in the 
(axial-) vector correlator, beginning with the finite temperature sector.   

\subsection{Hadronic Approaches I: Finite Temperature}
\label{sec_temp}
\begin{figure}[!htb]
\bce
\epsfig{file=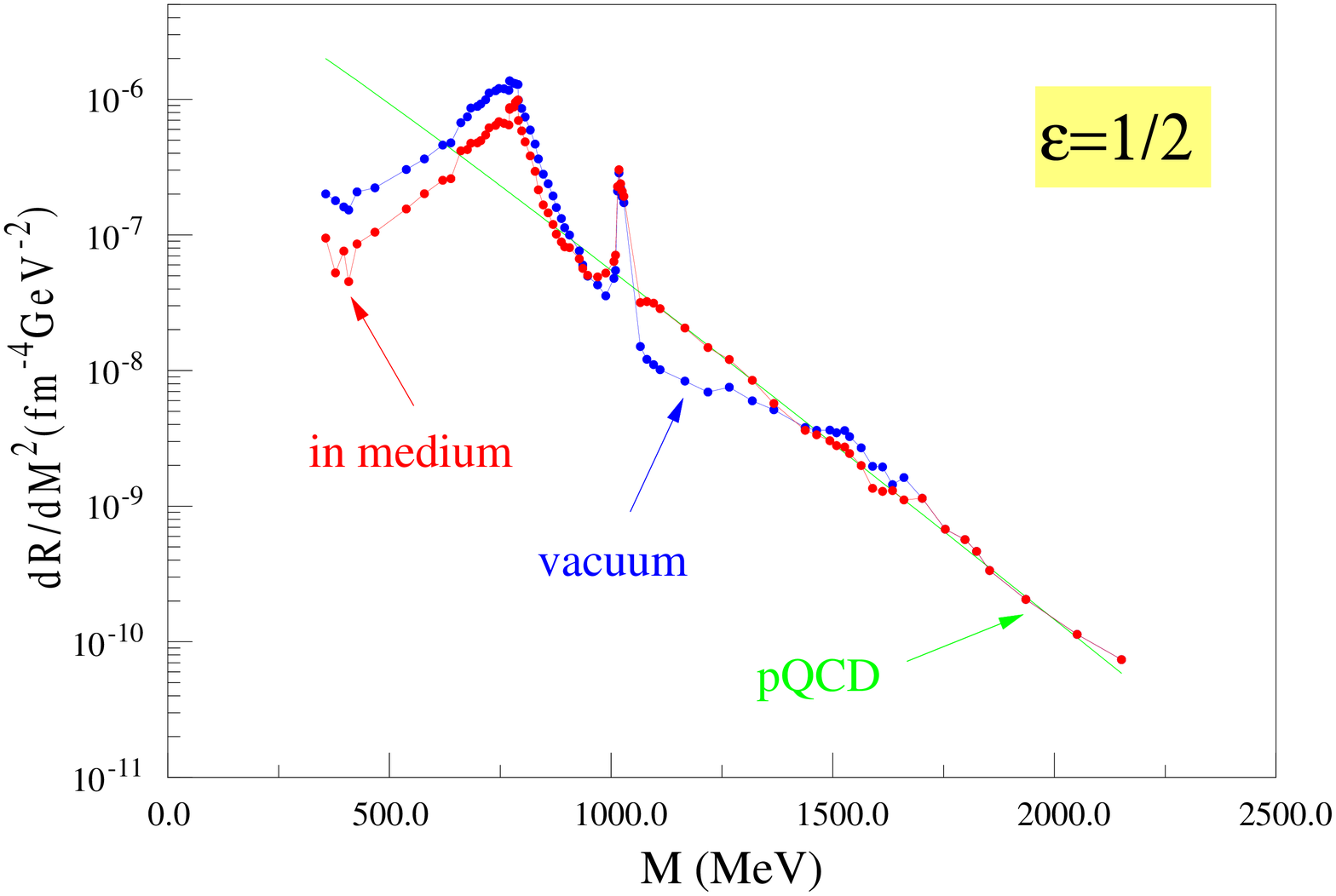,width=11.5cm}
\ece
\vspace{-1.5cm}
\caption{Three-momentum integrated dilepton production rates in hot 
baryon-free matter at temperature $T=160$~MeV using the free 
electromagnetic correlator (dark dots labelled 'vacuum'), 
the fully mixed one from Eq.~(\ref{vamix}) (light dots labelled 
'in-medium')~\protect\cite{Huang} and the perturbative QCD one 
from Eq.~(\protect\ref{ImPiem}) (dashed line, 'pQCD'). Note the close
agreement of the latter two between 1 and 1.5~GeV.} 
\label{fig_dlmix}
\end{figure}

Let us first concentrate on the model independent approaches that are
typically coupled with low temperature expansions. Dey \etal~\cite{DEI90} 
have shown that, in the chiral limit, the leading effect is a mere
mixing of vector and axialvector correlators with no medium effects
in the spectral shapes themselves, \ie,
\bea
\Pi^{\mu\nu}_V(q) &=& (1-\varepsilon) \ \Pi^{\circ\mu\nu}_V(q)
+\varepsilon \ \Pi^{\circ\mu\nu}_A(q)
\nonumber\\
\Pi^{\mu\nu}_A(q) &=& (1-\varepsilon) \ \Pi^{\circ\mu\nu}_A(q)
+\varepsilon \ \Pi^{\circ\mu\nu}_V(q) \ 
\label{vamix}
\eea
with $\varepsilon=T^2/6f_\pi^2$ and $\Pi^{\circ}$ denoting the vacuum 
correlators. When naively extrapolating to chiral restoration, 
where $\varepsilon=1/2$, one obtains $T_c^\chi=160$~MeV, close to what has
been found in lattice calulations. Even more surprising is the fact
that when calculating the three-momentum integrated dilepton production
rate, $dR/dM^2$, from $\Pi_V$ in Eq.~(\ref{vamix}), it coincides
with the result from perturbative $q\bar q$ annihilation 
starting from masses just beyond
the $\phi$ resonance, cf.~Fig.~\ref{fig_dlmix}. 
Thus, chiral restoration is coupled to a reduction of the duality threshold 
from 1.5 to $\sim$1~GeV, being a 'weak' temperature effect.  At the same
time, the light vector meson resonance structures are not affected.    

Another inherently model independent analysis has been pursued by 
Steele, Yamagi\-shi and Zahed (hereafter referred to as SYZ)~\cite{SYZ1}
within the so-called master formula framework. It uses  
chiral Ward identities based on broken chiral symmetry to express
correlators in terms of experimentally accessible (vacuum) on-shell 
scattering matrix elements in connection with a virial-type pion-density 
expansion. The vector correlator, \eg, takes the form
\bea
{\rm Im} \Pi_{V}^{\mu\nu} &=&\ {\rm Im} \Pi_V^{\circ\mu\nu}
+ \frac{1}{f_\pi^2} \int \frac{d^3k}{(2\pi)^3 2\omega_\pi(k)}  \
f^\pi(\omega_\pi(k);T) \ 
\langle \pi_{th} |j^\mu_{\rm em} j^\nu_{\rm em}|\pi_{th} \rangle 
\nonumber\\
\langle \pi_{th} |j^\mu_{\rm em} j^\nu_{\rm em}|\pi_{th} \rangle
&=& -4 \ {\rm Im} \Pi_V^{\circ\mu\nu}(q) +  
2 \ {\rm Im} \Pi_A^{\circ\mu\nu}(k+q) - 
2 \ {\rm Im} \Pi_A^{\circ\mu\nu}(k-q) + \cdots \ , 
\label{Pi_syz}
\eea
where the integration is over on-shell pion states from the heat bath.
The expansion parameter turns out to be $\kappa_\pi= n_\pi / 2m_\pi f_\pi^2$,
which is sufficiently small up to temperatures of about $T\simeq 140$~MeV. 
Clearly, Eq.~(\ref{Pi_syz}) also exhibits the $V$-$A$ mixing, which
is consistent with Eq.~(\ref{vamix}) in the chiral limit. 

Another class of approaches uses chiral Lagrangians to study the finite 
temperature behavior of (axial-) vector mesons. 
{\it E.g.}, employing the gauged linear $\sigma$ model
and imposing vector dominance, Pisarski found that in the vicinity of 
the phase transition point, a lowest order 
loop expansion gives  selfenergy corrections to $\rho$ and $a_1$ masses
such that~\cite{Pisa95}
\beq
m_\rho^2(T_c^\chi)=m_{a_1}^2(T_c^\chi)=\frac{1}{3} (2m_\rho^2+m_{a_1}^2) \ , 
\eeq
\ie, the masses merge in between their free values (with no dramatic
changes of the in-medium widths).    
Another variant is the 'Hidden Local Symmetry' scheme, which
in its minimal version does not include the $a_1$ meson. 
Nevertheless, the low-energy mixing theorem (\ref{vamix})
is satisfied through temperature corrections 
of the VDM coupling constant, $g_{\rho\gamma}(T)$~\cite{LSY95}.  
The application to calculating the in-medium pion electromagnetic
form factor within  a low-temperature pion-loop expansion
shows a strong reduction and broadening of the $\rho$ resonance
structure (with practically no mass shift), 
leaving a rather small enhancement in the corresponding
dilepton rates towards the two-pion threshold~\cite{SK96}.  

A third avenue of finite temperature calculations proceeds
with effective meson Lagrangians~\cite{GK91,GL94,Ha95,RCW,EIK,Gao99,RG99} 
where the emphasis is on incorporating all  phenomenologically 
important scattering processes which also go beyond $SU(2)$ chiral symmetry.  
This necessarily implies a less systematic implementation of the symmetry
properties, although the employed interaction vertices do satisfy basic 
requirements of gauge invariance and soft pion theorems. Concerning 
the in-medium $\rho$ spectral function, recent kinetic theory~\cite{Gao99}
as well as many-body-type calculations~\cite{RG99} reach quantitative 
consensus that in a meson gas at $T=150$~MeV 
the in-medium broadening amounts to $\sim$~80~MeV with 
insignificant mass shift.

\begin{figure}[!htb]
\vspace{-2cm}
\bce
\epsfig{file=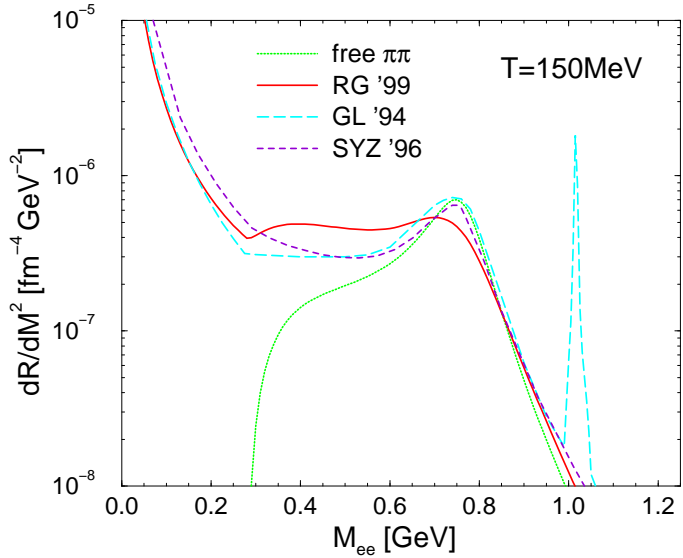,angle=-90,width=10cm}
\ece
\vspace{-1cm}
\caption{Three-momentum integrated dilepton production rates in hot
baryon-free matter at $T=150$~MeV in the hadronic approaches of
Gale/Lichard~\protect\cite{GL94} (long-dashed line),
Steele \etal~\protect\cite{SYZ1} (short-dashed line) and
Rapp/Gale~\protect\cite{RG99} (full line). The dotted line represents
free $\pi\pi$ annihilation.}
\label{fig_mesrates}
\end{figure}

Finally, we compare in 
Fig.~\ref{fig_mesrates} three different calculations 
of dilepton rates in thermal meson matter, the chiral reduction
approach (SYZ), many-body spectral function calculations (RG)
and an incoherent summation of individual decay rates using kinetic
theory (GL). 
All results agree that there is a moderate (factor 3-5) 
enhancement over free
$\pi\pi$ annihilation for invariant masses below the free $\rho$ 
(the somewhat larger excess from Ref.~\cite{RG99} being mostly
due to Bose-enhancement factors in the $\rho\to\pi\pi$ width not
included in the other two curves). In addition the thermal broadening
in the spectral function approach entails a $\sim$~30\% 
suppression in the $\rho$ resonance region. 

\subsection{Hadronic Approaches II: Finite Density}
\label{sec_dens}
The most famous approach, which integrally fueled the vigorous 
activity in the field, is the mean-field based analysis of Brown and
Rho (BR)~\cite{BR91} using arguments of scale invariance of the QCD 
Lagrangian. It culminated in the so-called BR-scaling conjecture
according to which all hadron masses (except for the Goldstone ones)
follow a universal density dependence linked to order parameters
of chiral restoration, $f_\pi$ or the quark condensate, as
\beq
\frac{\chi^*}{\chi_0}=\frac{f_\pi^*}{f_\pi}=\frac{m_\sigma^*}{m_\sigma}=
\frac{m_N^*}{m_N} = \frac{m_\rho^*}{m_\rho} = \frac{m_\omega^*}{m_\omega}
 \ , 
\eeq 
where quantities with an asterisk refer to the in-medium values. 
$\chi$ denotes the soft component of the (scalar) glueball field 
which has been introduced
on the effective chiral Lagrangian level to incorporate  
the same scaling properties as in QCD. Its vacuum expectation value
$\chi_0$ is associated with the soft part of the gluon condensate 
that is actually melted
in the chiral transition, being realized by the vanishing
of all hadron masses. This hypothesis has been successfully applied in
describing the low-mass dilepton enhancement at the 
CERN-SpS~\cite{CEK,LKBS}. 

The chiral reduction scheme mentioned above  
has also been extended to include nucleons~\cite{SYZ2}.
The medium effects in the vector correlator have been inferred as
\beq
\Pi_V^{\mu\nu} = \Pi_V^{\circ\mu\nu}
+  \int \frac{d^3p}{(2\pi)^3 2E_N(p)} \
f^N(E_N(p);\mu_N,T)  \ \langle N|j^\mu_{\rm em}|\alpha\rangle \ 
\langle\alpha|j^\nu_{\rm em} |N\rangle 
\eeq
through empirical information on the photon Compton tensor on the
nucleon with intermediate states 
$|\alpha\rangle= |\pi N\rangle, |\Delta(1232)\rangle,  |N(1520)\rangle$.   
Similar to other approaches discussed below, nucleons have been found
to impact the correlator stronger than thermal pions. 

Many finite density investigations have been carried out within (chiral)
effective Lagrangian frameworks. The early works~\cite{HFN92,CS92}
have mainly focused on modifications in the pion cloud of the $\rho$
through  nucleon- and delta-hole 
excitations well-known from nuclear optical potentials. 
However, model constraints imposed from nucleon/nuclear photoabsorption data
as well as $\pi N \to \rho N$ scattering data~\cite{KKW97,UBRW,RUBW,Fr98}
enforced the use of rather soft $\pi NN$ and 
$\pi N\Delta$ vertex form factors, suppressing the nuclear effects
in the pion cloud of the $\rho$.  
A more important role seems to be plaid by direct $\rho N$ scattering
into baryonic resonances (the so-called 'Rhosobars') as first proposed
by Friman/Pirner~\cite{FP97} for the $P$-wave (parity '+') 
states $N(1720)$ and $\Delta(1905)$. Also in this context, 
nucleon/nuclear photoabsortpion data
provided valuable information marking the $S$-wave (parity '--') 
states, in particular the $N(1520)$, as most relevant~\cite{PPLLM,RUBW}
 when moving into the timelike dilepton regime. 
The generic result of these calculations is that the vector
meson spectral functions undergo a strong broadening in cold nuclear 
matter. For the $\rho$ meson at normal nuclear matter density 
($\varrho_0=0.16$~fm$^{-3}$) it amounts to 200--300~MeV, which is 
a factor of 2--3 larger than at comparable
densities in a pure meson gas. 

\begin{figure}[!htb]
\begin{minipage}[t]{7.5cm}
\epsfig{file=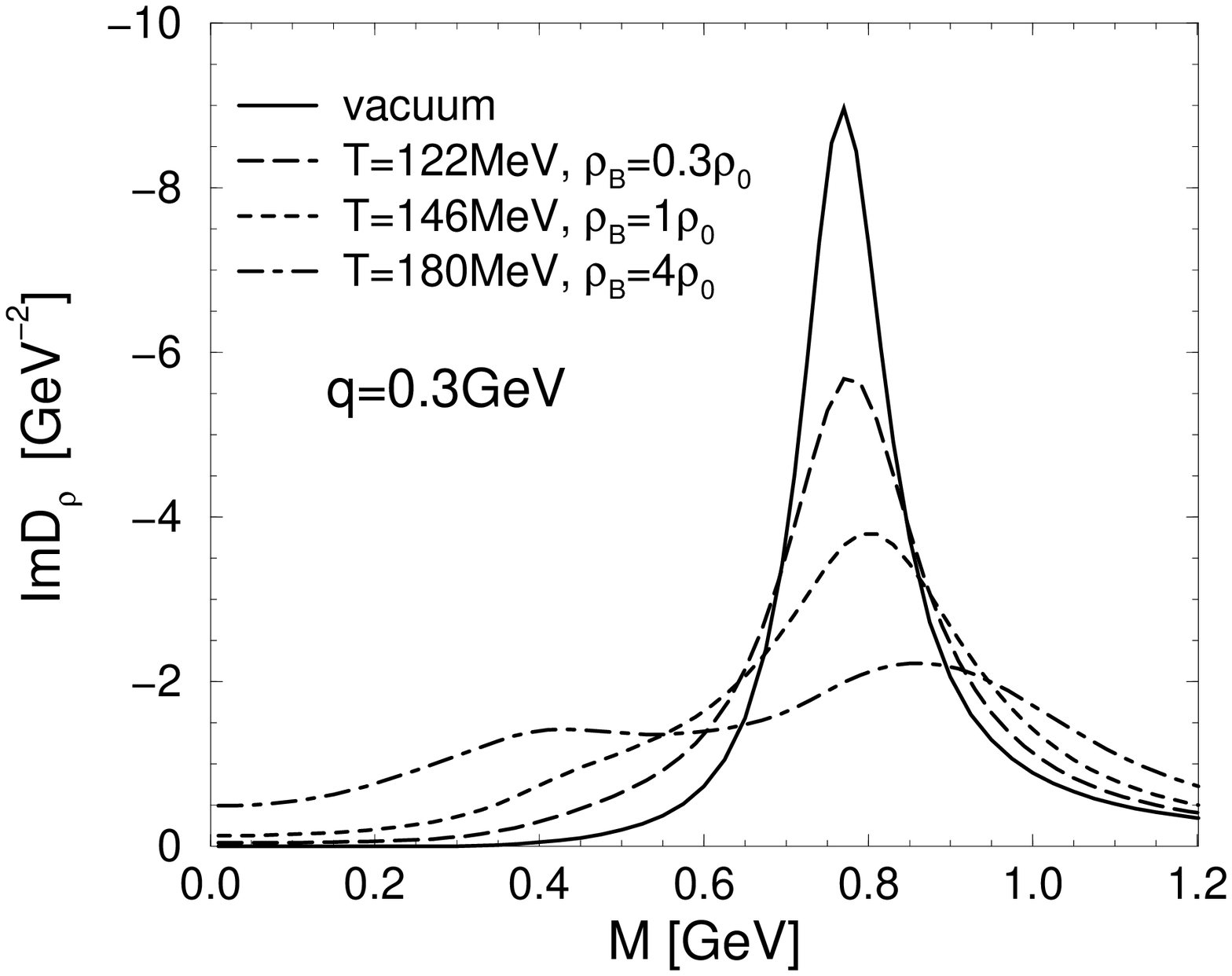,width=7.5cm}
\vspace{-1.5cm}
\caption{$\rho$ spectral function in hot hadronic matter at fixed
baryon chemical potential $\mu_B=408$~MeV.}
\label{fig_Arho}
\end{minipage}\hfill
\begin{minipage}[t]{7.5cm}
\epsfig{file=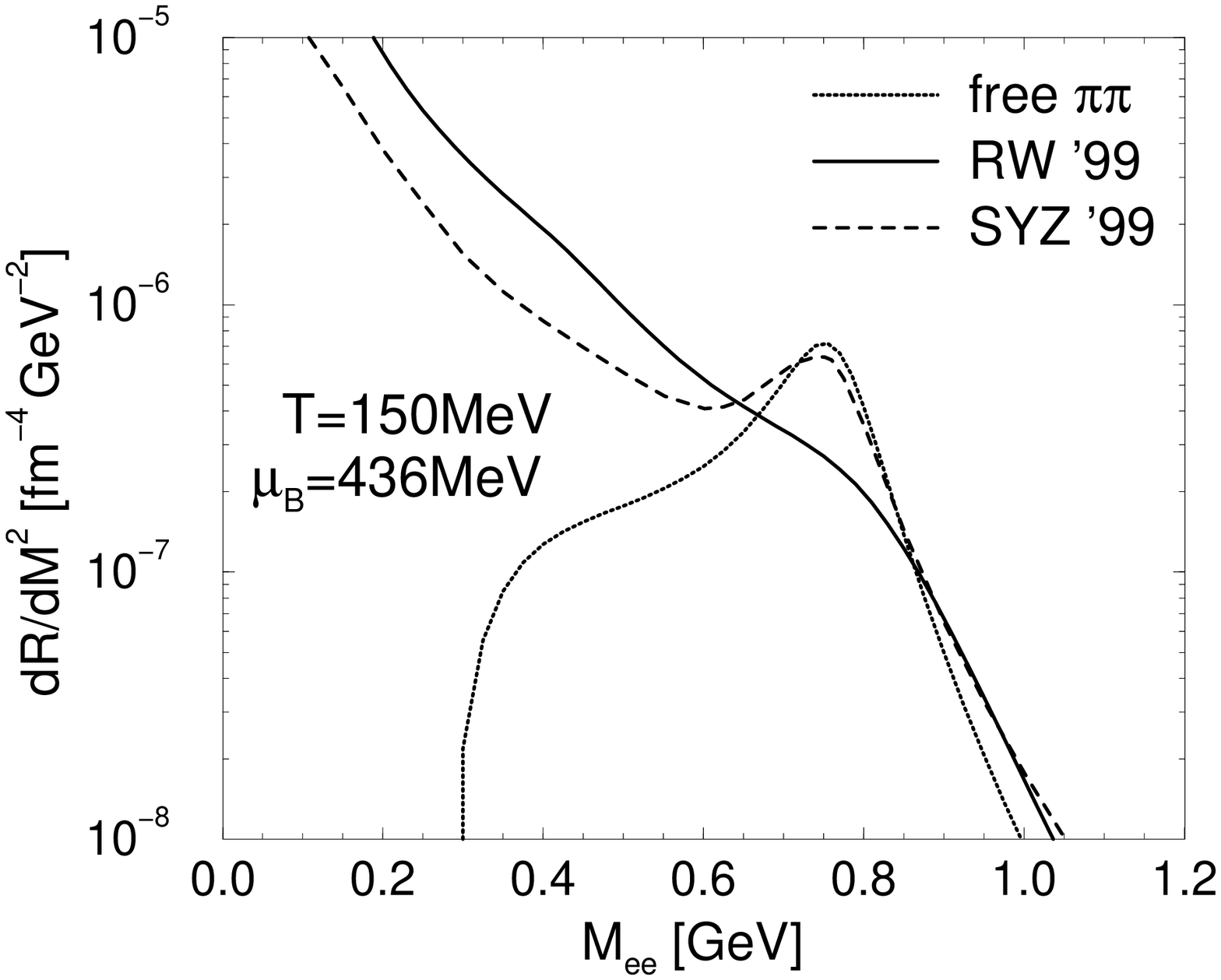,width=7.5cm}
\vspace{-1.5cm}
\caption{Dilepton rates in hot and dense matter ($\varrho_B$=1.5$\varrho_0$) 
within the spectral function~\protect\cite{RW99} (solid line)
and the chiral reduction 
approach~\protect\cite{SYZ2}(dashed line).}
\label{fig_dlrates}
\end{minipage}
\end{figure}

An important theoretical consistency check on the phenomenological
models can be inferred from QCD sum rule analyses. In Ref.~\cite{LPM}
a simple Breit-Wigner parametrization for the $\rho$ spectral function with
variable width and mass has been injected into the phenomenological side
to search for values that satisfy the sum rule. It turns out that there 
\begin{figure}[!htb]
\begin{minipage}[t]{7cm}
\bce
\vspace{-0.5cm}
\epsfig{file=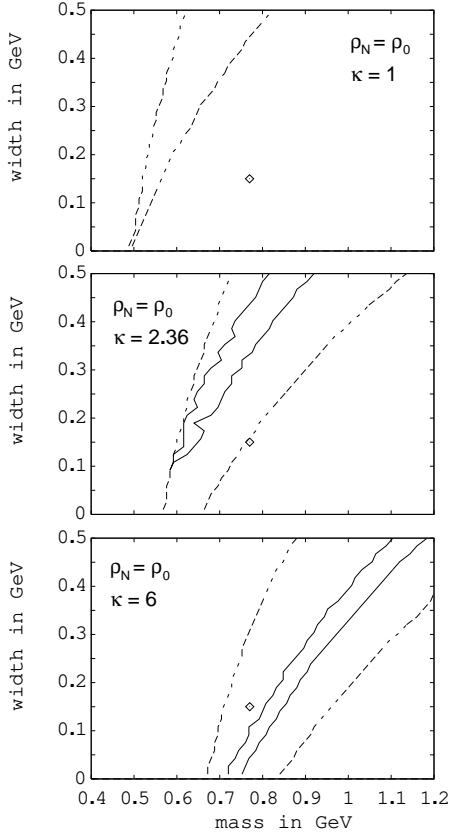,width=6cm}
\ece
\vspace{-1cm}
\caption{QCD sum rule allowed bands for in-medium $\rho$-mass and 
width~\protect\cite{LPM}.}
\label{fig_qcdsr}
\end{minipage}\hfill
\begin{minipage}[t]{8cm}
is in fact no unique prediction but rather bands of allowed values
in the mass-width plane, see Fig.~\ref{fig_qcdsr} for nuclear density 
(dashed and solid lines border the regions where 
the difference between {\it l.h.s.}~and {\it r.h.s.}~of the sum rule 
is less than 1\% and 0.2\%,
respectively). The correlation is such that one either has small mass and 
width (consistent with an earlier analysis~\cite{HL92} predicting
an in-medium mass decrease), or both increasing being consistent with 
the hadronic model calculations discussed above
as demonstrated in Ref.~\cite{KKW97}.  
The detailed location of the allowed bands depends somewhat on the  
assumptions made about the density-dependence of the quark and gluon 
condensates entering the operator product expansion in the sum rule, in 
particular the not very well-known value 
of the four quark condensate encoded in the factorization 
parameter $\kappa=\langle \bar qq \bar qq\rangle /\langle \bar qq\rangle^2$.
In Ref.~\cite{KKW97} the value of 2.36 (middle panel in Fig.~\ref{fig_qcdsr})
was fixed by requiring an optimal fit to the vacuum spectrum.  

Combining  the finite temperature (mesonic) 
and density (baryonic) effects through the
\end{minipage}
\vspace{-1.cm}
\end{figure}

\noindent
various selfenergy contributions in the $\rho$ propagator, 
\beq
D_\rho(M,q;\mu_B,T)=[M^2-(m_\rho^{(0)})^2-\Sigma_{\rho\pi\pi}-
\Sigma_{\rho M}-\Sigma_{\rho B}]^{-1} \ , 
\eeq 
leads to typical results as shown in Fig.~\ref{fig_Arho}~\cite{RW99}. 
The contributions to the broadening in the imaginary part are due to 
$\sim$~30\% mesonic and $\sim$~70\% baryonic effects,  in particular
the $S$-wave $\rho N$ resonances. At the highest temperature/density 
the second maximum structure around $M\simeq 400$~MeV 
is indeed due to the $N(1520)$,
which, in a selfconsistent treatment~\cite{PPLLM}, builds up a large 
in-medium width itself. The real
part of $D_\rho$ (not shown) 
becomes very flat making the concept of a mass (defined
by its zero-crossing) meaningless. 

Corresponding dilepton production rates from hot and dense hadronic matter
are compared with results from the chiral reduction scheme in 
Fig.~\ref{fig_dlrates}. 
Both approaches agree on a substantial, baryon-dominated enhancement
below the free $\rho$ mass (the quantitative differences can
be further traced down to different  relative strengths in various 
subprocesses; in particular, the N(1520) contribution, being
determined by photoabsorption spectra, is stronger in the RW calculations),
 but differ {\em qualitatively} in the 
$\rho$ resonance region. This can be understood as follows: 
The SYZ curve, being based on a virial-type expansion, is essentially 
proportional to the density of the surrounding matter, 
$dR/dM^2\propto (C_\pi n_\pi + C_B \varrho_B)$ with some coefficients
$C_\pi$, $C_B$.  The spectral function results behave as 
\beq
\frac{dR}{dM^2}\propto {\rm Im} D_\rho \propto \left\{
\begin{array}{lll}
{\rm Im} \Sigma_\rho / m_\rho^4 & \propto \ (\tilde C_\pi n_\pi + 
\tilde C_B \varrho_B) 
& , \ m_\rho^2\gg M^2, {\rm Im} \Sigma_\rho 
\\
1/{\rm Im} \Sigma_\rho & \propto \ 1/(\tilde C_\pi n_\pi + 
\tilde C_B \varrho_B) 
& , \ m_\rho \simeq M \ , 
\end{array} \right . 
\eeq
\ie, parametrically identical to the SYZ rates at low mass, but,
as a consequence of the resummations in the propagator, 
proportional to the inverse densities in the resonance region. 
\begin{figure}[!htb]
\vspace{-1cm}
\bce
\epsfig{file=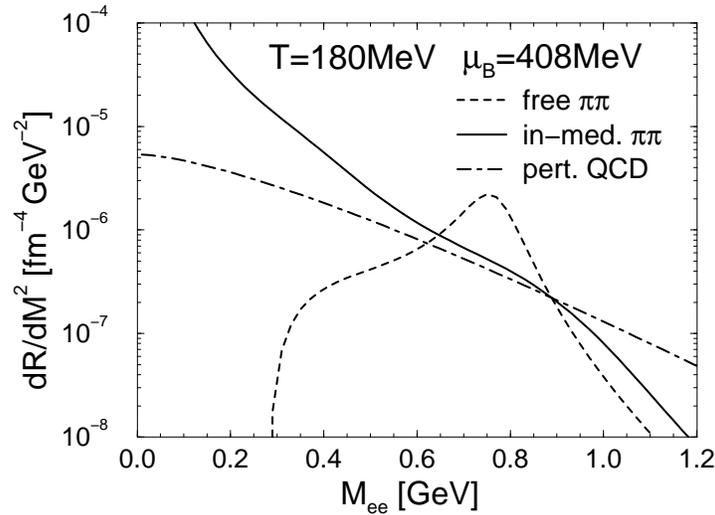,width=10cm}
\ece
\vspace{-1.5cm}
\caption{In-medium hadronic versus perturbative $q\bar q$ 
dilepton production rates.}
\label{fig_dual2}
\end{figure}
This strong smearing provokes yet another comparison to the perturbative 
$q \bar q$ annihilation rates, displayed in Fig.~\ref{fig_dual2}:
at extreme conditions the hadronic and the partonic description 
indeed coincide rather well down to invariant masses
of about 0.5~GeV (at masses above 1~GeV, the hadronic vector correlator,
being saturated by the $\rho$ meson, lacks the contributions from 
4-pion states etc.;  the agreement at very low masses might improve  
once soft (Bremsstrahlung-type) $\alpha_s$ corrections
are included). Although thermal loop corrections to the partonic rates
at small masses are not yet well under control, Fig.~\ref{fig_dual2}
seems to indicate that the duality threshold is further reduced to well
below 1~GeV, this time as a consequence of 'strong' 
(predominantly baryon-driven) resummation effects.

\section{Dilepton Spectra at CERN Energies}
\label{sec_dlspec}
\begin{figure}[!htb]
\begin{minipage}{8.7cm}
\epsfig{file=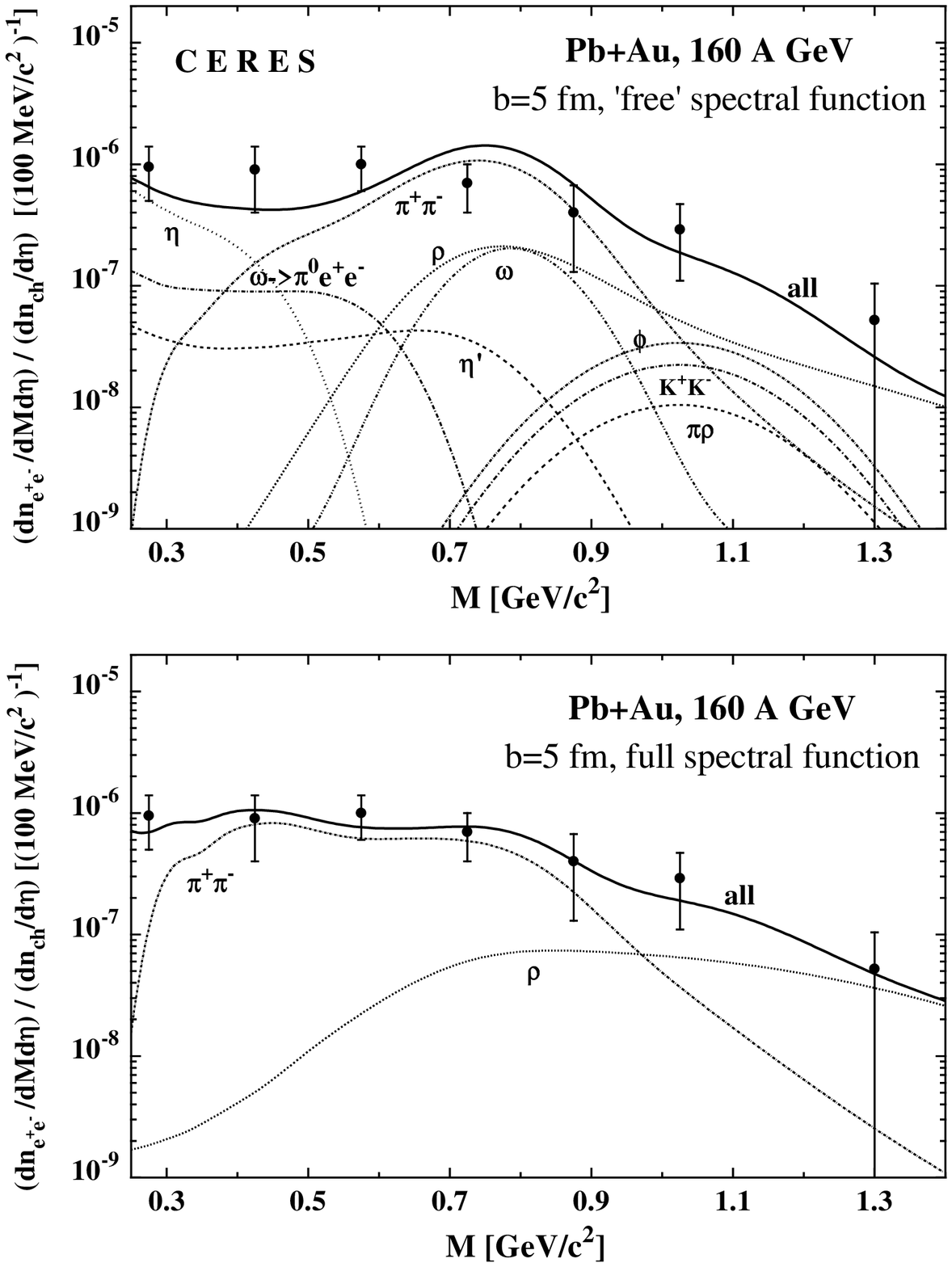,width=8.7cm}
\end{minipage}
\begin{minipage}{6cm}
\epsfig{file=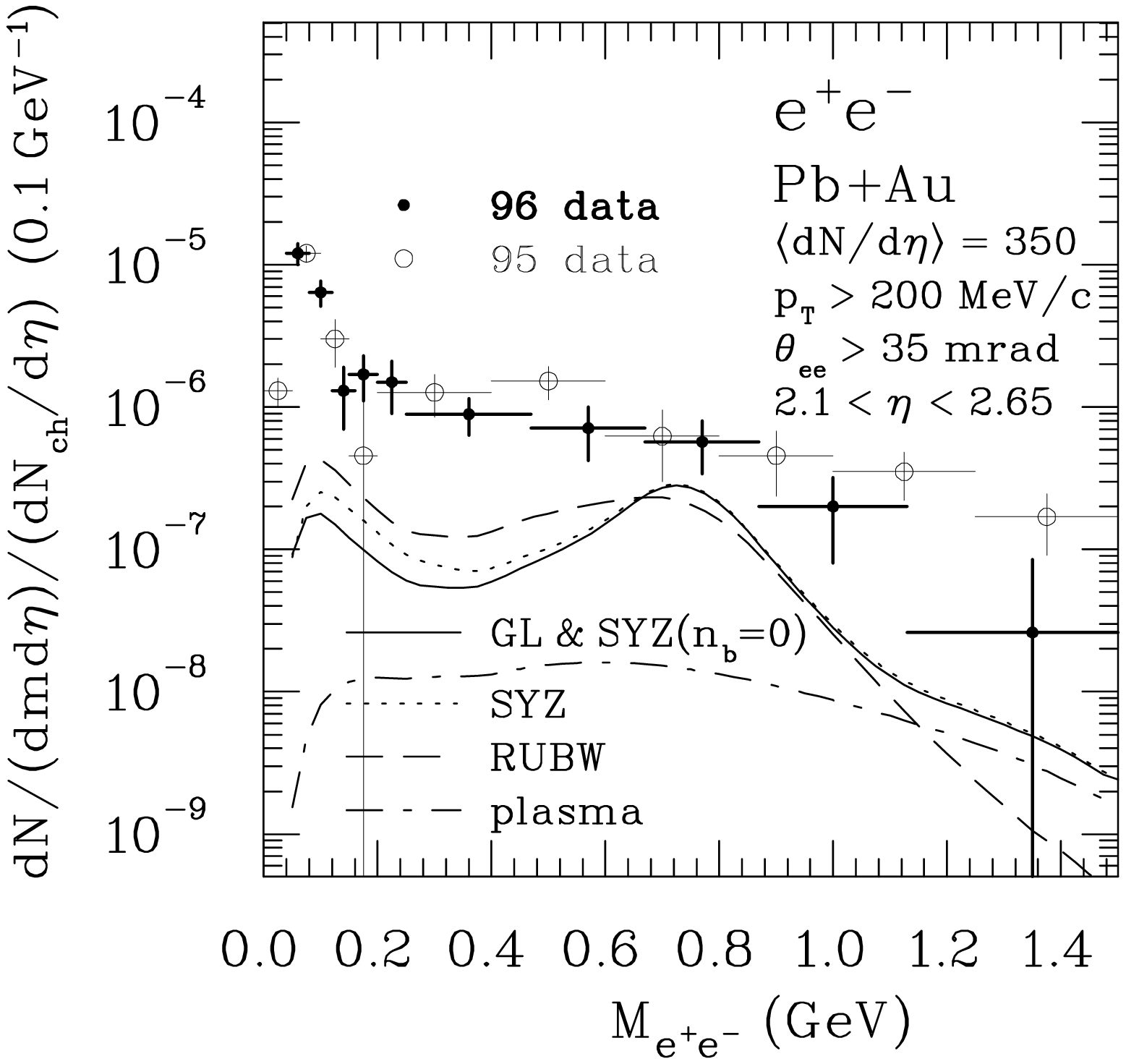,width=6cm}
\epsfig{file=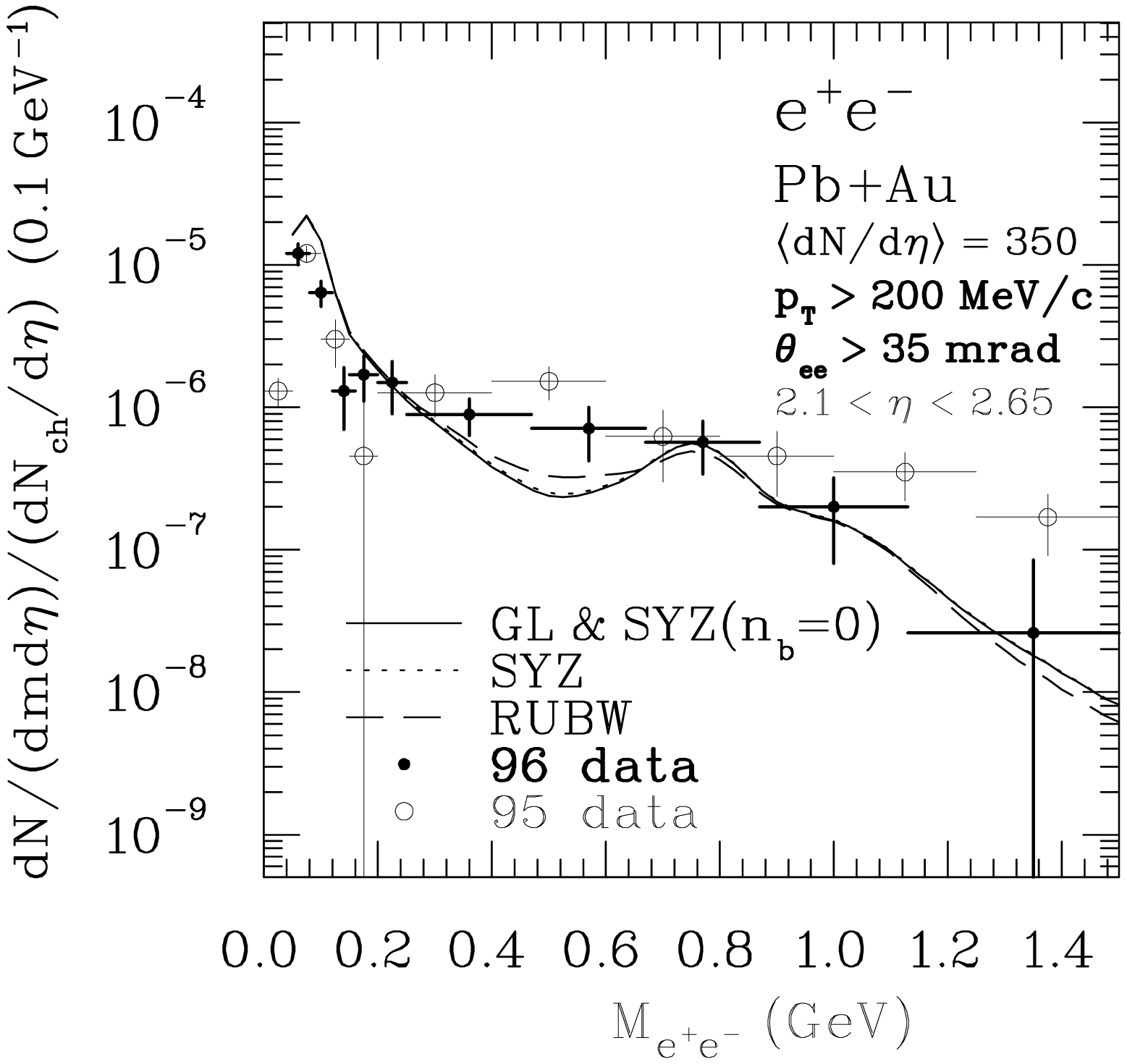,width=6cm}
\end{minipage}
\caption{CERES dilepton data~\protect\cite{ceres} 
compared to HSD transport~\protect\cite{CBRW} 
(left panel) and hydrodynamic~\protect\cite{HP99} (right panel)
simulations.}
\label{fig_dlcomp}
\end{figure}

\begin{figure}[!htb]
\begin{minipage}[t]{7.3cm}
\epsfig{file=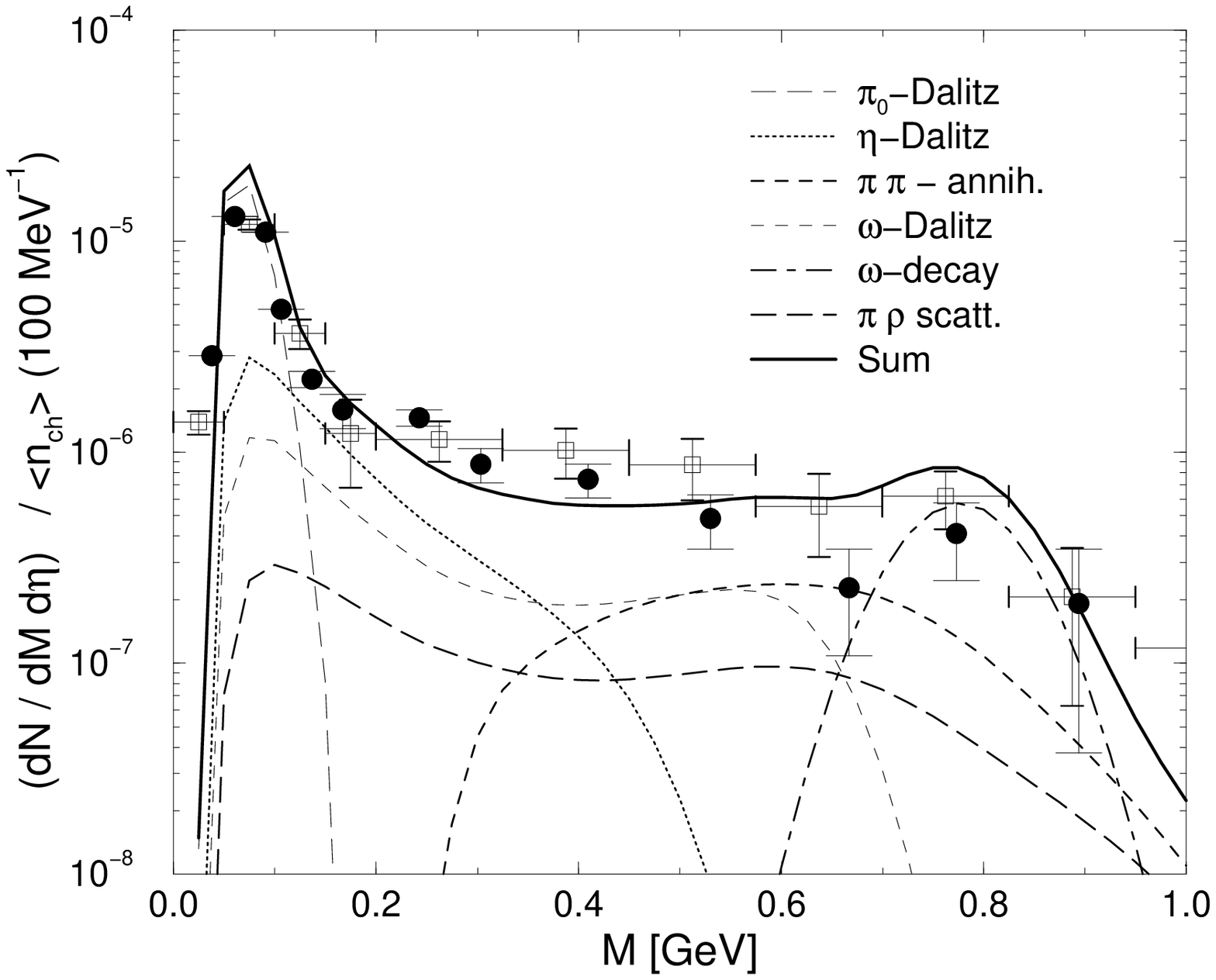,width=7.3cm}
\vspace{-1.5cm}
\caption{CERES dilepton data~\protect\cite{ceres} compared to RBUU transport 
results~\protect\cite{Koch99}.}  
\label{fig_dlkoch}
\end{minipage}\hfill
\begin{minipage}[t]{8.1cm}
\epsfig{file=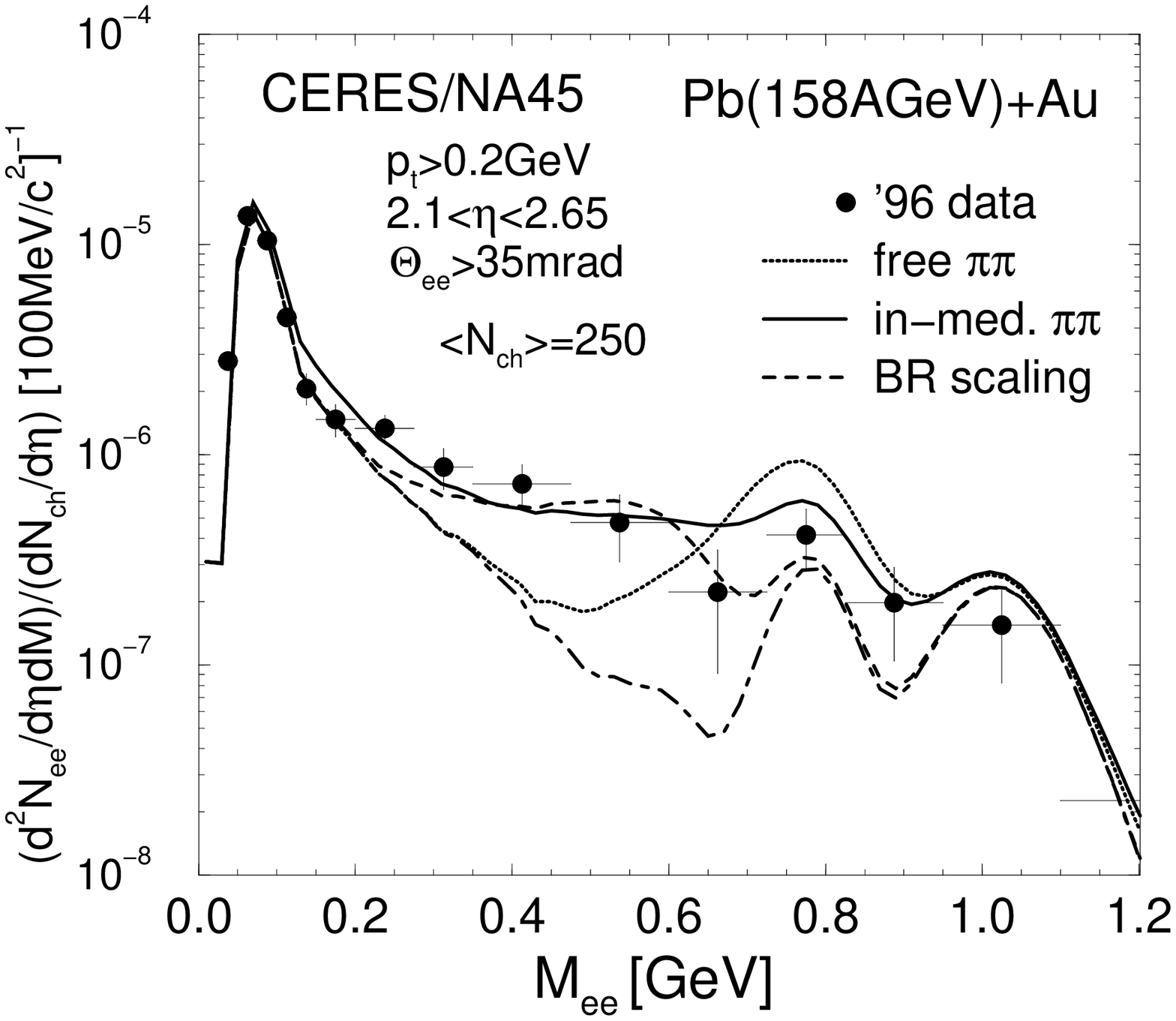,width=8.1cm}
\vspace{-1.5cm}
\caption{CERES dilepton~\protect\cite{ceres} data compared to thermal 
fireball calculations~\protect\cite{RW99}.} 
\label{fig_dlrapp}
\end{minipage}
\end{figure}

An evaluation of dilepton spectra in URHIC's requires the convolution
of the elementary production rates (processes) over the space-time
evolution of the hadronic fireball. The additional contribution
from  electromagnetic decays of hadrons after freezeout
can be rather reliably assessed once the final state hadron
abundancies are known (which is the case for $\pi^0$'s and $\eta$'s,
but difficult for $\omega$'s).
Both microscopic transport and hydrodynamic simulations have been
proven successful in describing the measured hadron spectra.
However, as illustrated in Fig.~\ref{fig_dlcomp}, they may differ substantially
in their prediction for in-medium produced dileptons, $N_{ee}^{med}$.
Irrespective
of whether the free or an in-medium $\rho$ spectral distribution
is employed, the transport calculations give a factor of $\sim$3
larger yields. Whereas the final pion number (fixed by experiment) 
is schematically given
by density times fireball volume, $N_\pi\propto n_\pi V_{FB}$, 
in-medium dilepton radiation (arising mostly from $\pi\pi$ annihilation)
behaves like $N_{ee}^{med}\propto n_\pi^2 V_{FB}$.
The discrepancy in the latter might point at an off-equilibrium
occupation of pions, \ie, finite pion-chemical potential, not included
in the hydro results, which as consequence do not describe the CERES
data even with an in-medium spectral function.      
On the other hand, the transport results of Koch~\cite{Koch99}
(Fig.~\ref{fig_dlkoch})  
come rather close to the low-mass enhancement around 0.4~GeV 
with only a rather moderate in-medium contributions. 
This is achieved
through a rather large $\omega$ Dalitz decay yield, which, however,
tends to imply an overestimation of the data at 0.8~GeV due
to direct $\omega\to ee$ decays. The upcoming improved mass 
resolution/statistics measurements from CERES will be essential
to clarify this issue. 
Fig.~\ref{fig_dlrapp} shows thermal fireball calculations~\cite{RW99}
where the temperature/density evolution is consistent with the  
recently determined {\em chemical} freezeout at SpS energies~\cite{pbm98}. 
Using entropy and baryon number conservation, and further assuming
effective pion number conservation towards {\em thermal} freezeout
leads to the build-up of pion chemical potentials reaching almost
80~MeV. The resulting dilepton spectra (supplemented with the CERES
cocktail~\cite{cktl}(dashed-dotted line) for hadron decays after 
freezeout) employing either the in-medium
$\rho$ spectral function (solid line) or the dropping $\rho$ mass 
conjecture (dashed line) are in reasonable agreement with experiment,
which also holds for transverse momentum dependencies~\cite{RW99}.

Finally, let us briefly comment on the implications of thermally 
produced dileptons for the intermediate
mass region as covered, \eg, by the NA50 experiment~\cite{bordalo}.
Preliminary results of a recent calculation~\cite{RS99} using the same thermal
fireball model as in Fig.~\ref{fig_dlrapp} (and an approximate
NA50 acceptance) indicate that the factor of $\sim$3 enhancement observed 
in the data for $M\simeq 1.5-2.5$~GeV
can indeed be accounted for without having to invoke any 'anomalous'
open charm enhancement. 
Note that above $M=1.5$~GeV there is no 
longer an issue of medium effects as hadronic and $q\bar q$ rates are 'dual'
already in vacuum. 

\section{Conclusions}
\label{sec_concl}
The last few years have witnessed continuous progress in understanding
the in-medium properties of vector mesons and the pertinent dilepton 
production rates and spectra in URHIC's.  
It has also become clear that a profound discussion of chiral symmetry
restoration needs to involve the chiral partner of the vector correlator, 
\ie, the axialvector ($a_1$) channel. 
Low temperature theorems have shown that the leading temperature effect
is a mere mixing between the two through interactions with thermal 
pions. When extrapolated to the phase transition region, this 'soft'
temperature effect leads to an additional degeneration of hadronic
and perturbative $q\bar q$ dilepton production rates down to masses
of about 1~GeV. At lower masses a strong broadening of the 
$\rho$ resonance, driven by the resummation of baryon-dominated
in-medium interactions, seems to induce a further lowering of the 
'duality threshold' to about 0.5~GeV. Large in-medium widths of the 
$\rho$ are supported by most phenomenological calculations and are
also consistent with QCD sum rules. A more rigorous 
link to chiral restoration requires advanced investigations of 
the in-medium $a_1$ properties.  

We have furthermore shown that the broadening scenario is compatible
with low-mass dilepton observables at SpS energies when employing 
microscopic transport calculations or thermal fireball evolutions 
including the build-up of moderate pion chemical potentials. 
However, a conclusive discrimination of the in-medium contribution
in the experimental spectra can only be achieved
with upcoming improved mass resolution/statistics measurements,
which are essential to separate direct $\omega$ decays.
Also, the commissioned 40~GeV run at the SpS will be most valuable
for a more precise assessment of high baryon-density effects.  
  
\vspace{1cm}

\noindent
{\bf ACKNOWLEDGEMENT}

\vspace{0.3cm}

\noindent
It is a pleasure to thank J. Wambach for the fruitful collaboration
on the presented topic over the last years. I am also grateful
for many productive conversations and collaborations 
with my colleagues at Stony Brook, Darmstadt, Gie{\ss}en, Montreal  
and Heidelberg.


\end{document}